\documentclass[prb,twocolumn,showpacs,preprintnumbers,amsmath,amssymb]{revtex4}

\usepackage{graphicx}
\usepackage{dcolumn}
\usepackage{bm}
\usepackage{subfigure}

\begin{document}

\title{Resistance of multilayers with long length scale interfacial roughness}

\author{Jason Alicea\footnote{Present address: Department of Physics, 
University of California, Santa Barbara, California 93117}}
 \email{aliceaj@physics.ucsb.edu}
\author{Selman Hershfield}
 \email{selman@phys.ufl.edu}
\affiliation{Department of Physics and National High Magnetic Field Laboratory,
University of Florida, Gainesville, FL 32611-8440}

\date{\today}

\begin{abstract}
The resistance of multilayers with interface roughness on a length
scale which is large compared to the atomic spacing is computed in
several cases via the Boltzmann equation. This type of roughness
is common in magnetic multilayers. 
When the electronic mean free paths
are small compared to the layer thicknesses, the current flow is
non-uniform, and the resistance decreases in the
Current-Perpendicular-to-Plane (CPP) configuration and increases
in the Current-In-Plane (CIP) configuration. For mean free paths
much longer than the layer thicknesses, the current flow is
uniform, and the resistance increases in both the CPP and CIP
configurations due to enhanced surface scattering. In both the CPP
and CIP geometries, the giant magnetoresistance can be either
enhanced or reduced by the presence of long length scale interface
roughness depending on the parameters.  Finally, the changes in
the CPP and CIP resistivities due to increasing interface
roughness are estimated using experimentally determined parameters.
\end{abstract}


\maketitle

\section{Introduction}

The study of metallic multilayers has been a very active area of
research in recent years. Of particular interest are alternating
layers of ferromagnetic and paramagnetic metals called magnetic
multilayers. A relatively small magnetic field aligns the magnetic
moments in the different ferromagnetic layers, leading to a large
magnetoresistance, which is called the giant magnetoresistance
(GMR).\ \cite{discovery,CouplingDiscovery} 
The GMR has technological applications in magnetic read
heads, magnetic sensors, and magnetic memory devices.
Consequently, it has been studied extensively, including such
effects as bulk and interface scattering, magnetic and
non-magnetic scattering, Fermi wave vector mismatch and magnetic
coupling between the layers.
\cite{LevyReview,GijsBauerReview,Summary2,FertReview}

One aspect of metallic multilayers which has been less studied 
theoretically is
the effect of long length scale fluctuations of the layer
thicknesses and heights. In an ideal multilayer the interfaces
between the layers are perfect planes. The thickness of each layer
and the height of each layer above the substrate would be
constant. Obviously, this is not the case in any real system. In
addition to interdiffusion and other atomic scale disorder at the
interfaces, the actual thicknesses and/or heights of the layers
can vary on a length scale which is large on the atomic scale.
Indeed, these long length scale fluctuations seem to be the rule
rather than the exception.

Since the Fermi wave vectors for these metals are of order the
atomic spacing, interdiffusion and atomic scale disorder are
strong sources of scattering.  Fluctuations on a length scale of
10 or more atomic spacings should not appreciably effect the
surface scattering. 
Nonetheless, long length scale disorder can be
important. 
It has been demonstrated both experimentally \cite{90degree}
and theoretically \cite{Slonczewski}
that nonplanar interfaces can create new kinds of magnetic
coupling between the layers.  This tends to
reduce the GMR because the fraction of a sample that is
antiferromagnetically aligned at zero applied field is reduced.
Consequently, many of the experiments that study the role
of interfacial roughness rescale the magnetoresistance by the
fraction of the sample which is antiferromagnetically coupled.
This fraction can be determined experimentally from magnetization
measurements.

There are two common geometries used in studying the giant
magnetoresistance. The current flows parallel to the layers in the
Current-In-Plane (CIP) geometry and normal to the layers in the
Current-Perpendicular-to-Plane (CPP) geometry. 
The CIP geometry has been more widely studied than the 
CPP geometry because measurements in the CIP 
configuration are easier to achieve experimentally.
Simultaneous measurements of the CIP magnetoresistance and roughness
using low angle X-ray scattering have been made on samples with 
roughness that has been systematically varied by changing growth 
parameters and annealing.
One set of experiments on Fe/Cr multilayers finds that the change 
in the resistivity between low and high field, $\Delta \rho$, rescaled
by the antiferromagnetic fraction, $AFF$, increases with roughness.\
\cite{SchullerFirstRoughness,SchullerAnnealing,SchullerSputteringPressure}
Schad \emph{et al}.\ also simultaneously measured the surface roughness 
and the magnetoresistance of Fe/Cr multilayers.\
\cite{SchadAFF,BruynseraedeInterface,Bruynseraede2}
In one set of experiments they find
that $\Delta \rho/AFF$ decreases with the increasing 
fluctuations in the layer heights while
the saturation resistivity decreases.\
\cite{SchadAFF,BruynseraedeInterface}
In another set of experiments on Fe/Cr multilayers which are dominated by
surface scattering they find that the CIP magnetoresistance 
increases with 
interface fluctuations, which were determined by the 
ratio of the vertical roughness amplitude
to the lateral correlation length.\
\cite{Bruynseraede2}

The effect of long length scale roughness on the CIP magnetoresistance
has been studied theoretically 
by Barnas and Bruynseraede \cite{BarnasBruynseraede}
and by Levy \emph{et al}.\ \cite{Levy}
Barnas and Bruynseraede studied the scattering between quantum states in
different layers allowing for uncorrelated variations in the layer thicknesses.
They find that the magnetoresistance can either increase of decrease
with roughness depending on the parameters in the problem.  Indeed
there are even cases where the magnetoresistance decreases and then
increases with roughness.  Levy \emph{et al}.\ studied the magnetoresistance
of multilayers deposited on grooved substrates, \cite{OnoGrooved,GijsGrooved} 
which produce roughness
which is correlated between the layers.  Using a general linear response
approach they find that the roughness mixes the CIP and CPP geometries,
leading to what is called the CAP or Current-at-an-Angle-to-Plane 
configuration.
Since the CIP magnetoresistance is typically smaller than the CPP 
magnetoresistance, this would tend to increase the magnetoresistance
with roughness.

The difficulty in making CPP measurements is due to the large surface
area relative to the thickness of the samples.  This results in a
CPP resistance which is too small to measure using conventional 
techniques and also makes the measurements
susceptible to inhomogeneous current paths.  These difficulties have
been overcome by making small area pillar samples, \cite{Gijs}
by using superconducting leads, \cite{BassCPPDetailComp} and 
by combinations of these two techniques.\ \cite{Schuller1}
Experiments on Co/Ag multilayers find that the CPP magnetoresistance decreases
with interface disorder while the CIP magnetoresistance increases.
\cite{PrattPressure}
Cyrille \emph{et al}.\ measured the CPP magnetoresistance of Fe/Cr
multilayers and quantified the roughness in their samples using 
both low angle X-ray
scattering and transmission electron micrographs of cross-sectional
samples.\  
\cite{Schuller1,Schuller2}
They find that $\Delta \rho/AFF$ increases with roughness
proportionally to the RMS deviation in the layer heights.
Still more recent experiments by Zambano \emph{et al}.\ see no change in 
the CPP magnetoresistance of Fe/Cr multilayers with increasing roughness.\
\cite{SeriesResistor}

The effect of long length scale interface fluctuations is
therefore unclear, with some work pointing to an increase in the
magnetoresistance, some a decrease, and some no change at all.
Some of the differences between the experiments is due to changing
a growth parameter like the sputtering pressure probably changes
more than just the interfacial roughness.  A theoretical calculation
can vary just the interface fluctuations and hence hope to isolate
the effect of long length scale interface fluctuations on the
magnetoresistance.
While the earlier theoretical work is consistent with the
CIP magnetoresistance experiments, the CPP magnetoresistance 
experiments remain unexplained.  In particular the extensive work
of Cyrille \emph{et al}.\ \cite{Schuller1,Schuller2}
which shows an increase in the magnetoresistance
with interface roughness is not possible to explain as a mixing
of the CPP and CIP geometries since the CIP magnetoresistance is
usually lower than the CPP one, as is the case in their measurements.

In this paper we examine the effects of long length scale
fluctuations in the layer thicknesses and heights using the
Boltzmann equation. A semiclassical approach like the Boltzmann
equation is a good choice for this problem because the length
scales involved are large compared to the atomic spacing.  Various
versions of the Boltzmann equation have also been used extensively
in modeling the GMR.\ \cite{BoltzmannTheory,ValetFert,ButlerBoltzmann,
StilesBoltzmann} 
The Boltzmann equation we use here has a simple, but current
conserving form for the scattering term, and we can solve it
essentially exactly numerically in the limits of long and short
mean free paths. Here, a long mean free path is much larger than
the layer thicknesses, and a short mean free path is much smaller
than the layer thicknesses. We consider both the CPP and CIP
geometries and find that the GMR can either increase or decrease
with interface roughness depending on the parameters in the
problem. Explicit predictions are made for when the GMR
increases and when it decreases.

The remainder of the paper is organized as follows. In Sect.\ II
we provide a formal solution to the Boltzmann equation which is
valid for arbitrary mean free paths.  Next, in Sect.\ III we apply
this formal solution to reproduce the well known results for the
CIP and CPP geometries when the interfaces are flat. In Sect.\ IV
we consider sinusoidal interfaces, which model the long length scale
interface fluctuations.  Both the long mean free path case (Sect.\
IV A) and the short mean free path case (Sect.\ IV B) are
considered. These two cases are compared numerically in Sect.\ IV
C. In Sect.\ V we use the results from Sect.\ IV to determine the
effect of long length scale interface disorder on the giant
magnetoresistance in the long mean free path case (Sect.\ V A) and
the short mean free path case (Sect.\ V B). In Sect.\ V C the
changes in the CPP and CIP resistivities due to increasing long
length scale roughness are estimated in the long and short mean
free path cases using experimentally determined parameters. All
the results are summarized in the conclusion, Sect.\ VI.

\section{Formalism}

\begin{figure}
  {\resizebox{8cm}{!}{\includegraphics{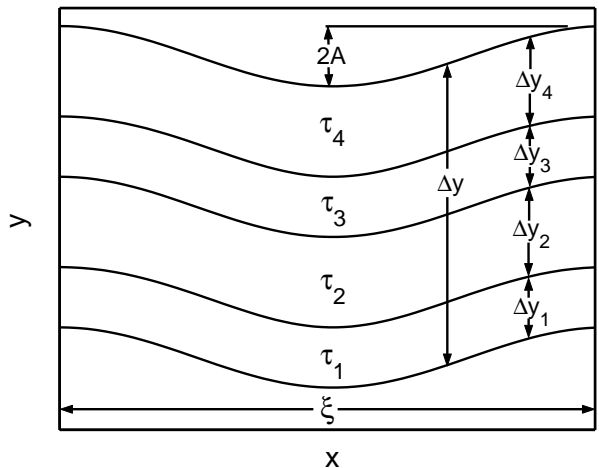}}}
  \caption{\label{fig:parameters} Schematic of a four-layer repeat unit
  with interfacial roughness. The interfaces are modeled as sine waves of
  amplitude $A$ and period
  $\xi$. Layer $i$ has a thickness $\Delta y_i$ and 
  uniform relaxation time $\tau_i$, and the
  total thickness of the repeat unit is $\Delta y$.
  We calculate the current density and conductivity due
  to roughness for currents flowing in the $y$-direction
  (CPP configuration) and $x$-direction (CIP configuration)
  for mean free paths which are long and short compared to $\Delta y$.}
\end{figure}

The model system we study is shown in Fig.\ \ref{fig:parameters}.
Multilayers of thickness $\Delta y$ are separated by sinusoidal
interfaces with amplitude $A$ and period $\xi$. Within each layer
the relaxation time, $\tau _i$, is constant. For numerical
considerations, our calculations are performed in two dimensions,
namely the multilayers are strips in a two-dimensional plane;
however, we nonetheless expect the results to remain qualitatively
the same when generalized to three dimensions. Moreover, in both
the long and short mean free path limits we will present analytic
expressions for our results, which have direct generalizations to
the three-dimensional case.

The Boltzmann equation we use represents elastic s-wave scattering
within a current-conserving right-hand side,
\begin{equation}
    {\bf v \cdot \nabla}_rf-e{\bf E \cdot \nabla}_pf =
    -\bigg{(}\frac{f-\overline f}{\tau}\bigg{)},
    \label{eq:Boltz}
\end{equation}
where $f=f(\bf r, \bf p)$ is the distribution function,
${\overline f} = {\overline f}(\bf r,|\bf p|)$ is the spherical
average of $f$ in momentum space, and $\tau = \tau(\bf r)$ is the
relaxation time.
To get the linear response conductivity the distribution function
is expanded to first order in the applied electric field,
    $f = f_{eq}+\delta f$,
where $f_{eq}$ is the equilibrium Fermi-Dirac distribution function
and $\delta f$ is proportional to the applied electric field, ${\bf E}$.
The linear response Boltzmann equation is
\begin{equation}
    {\bf v \cdot \nabla}_r\delta f-e{\bf E \cdot v}
\left( \frac{\partial f_{eq}}{\partial \epsilon } \right) =
    -\bigg{(}\frac{\delta f-\overline {\delta f}}{\tau}\bigg{)}.
    \label{eq:Boltz2}
\end{equation}

This equation may be further simplified using the fact that the
temperature is far below the Fermi temperature in these metals.
Consequently, the energy derivative of the equilibrium Fermi
function in Eq.\ (\ref{eq:Boltz2}) is approximately a delta
function which pins the energy to the Fermi energy. One can define
another distribution function on the Fermi surface, $g$,
\begin{equation}
    \delta f = \bigg{(}-\frac{\partial f_{eq}}{\partial \epsilon}\bigg{)}g
    \label{eq:g}
\end{equation}
so that the Boltzmann equation becomes
\begin{equation}
    {\bf v \cdot \nabla}_r g + e{\bf E \cdot v} =
    -\bigg{(}\frac{g-\overline {g}}{\tau}\bigg{)}.
    \label{eq:Boltz4}
\end{equation}

This first order differential equation can be integrated
to find $g$ and hence $f$.  As a first step the deviation
of $g$ from its spherical average $\overline{g}$ is defined as
$\delta g$,
\begin{equation}
    \delta g = g-\overline {g}.
    \label{eq:deltag}
\end{equation}
Substituting $\delta g$ into Eq.\ (\ref{eq:Boltz4}), the Boltzmann
equation on the Fermi surface is now
\begin{equation}
    {\bf v \cdot \nabla}_r \delta g + \frac{\delta g}{\tau} =
    -e{\bf E \cdot v} - {\bf v \cdot \nabla}_r \overline {g}.
    \label{eq:Boltz5}
\end{equation}
Both the gradient of the electrical potential, ${\bf E} = - {\bf
\nabla}\phi$, and the gradient of $\overline{g}$ appear on the
right hand side of Eq.\ (\ref{eq:Boltz5}).  Consequently, it is
useful to define the electro-chemical potential, $V$,
\begin{equation}
    V = \phi - \frac{\overline g}{e},
    \label{eq:V}
\end{equation}
so that there is only a single gradient on the right-hand side,
\begin{equation}
    {\bf v \cdot \nabla}_r \delta g + \frac{\delta g}{\tau} =
 e{\bf v \cdot \nabla}V .
    \label{eq:Boltz5b}
\end{equation}

Equation (\ref{eq:Boltz5b}) can be solved by integrating along
paths in phase space. Letting $({\bf r}(s),{\bf p}(s))$ be a
trajectory in phase space which satisfies the equation of motion,
${\bf \dot{r}} = {\bf v}$ and ${\bf \dot{p}} = 0$, the
distribution function along this path is
\begin{equation}
   {\tilde g}(s) = \delta g({\bf r}(s),{\bf p}(s)) .
   \label{eq:gtilde}
\end{equation}
Substituting Eq.\ (\ref{eq:gtilde}) into Eq.\ (\ref{eq:Boltz5b}),
one obtains a first order ordinary differential equation,
\begin{equation}
    \frac{d{\tilde g}}{ds} +\frac{{\tilde g}}{\tau} = e\frac{dV}{ds}.
    \label{eq:Boltzlast}
\end{equation}
The general solution of this equation is
\begin{eqnarray}
    \tilde g(s_f) &=& \exp\left(
    -\int _{s_i}^{s_f} \frac 1{\tau (s)}ds
    \right)\tilde g(s_i)
    \label{eq:gensol}
    \\
    &+&\int_{s_i}^{s_f}\exp\left(
    -\int _{s}^{s_f} \frac 1{\tau (s')}ds'
    \right) e\frac{dV}{ds}ds,
    \nonumber
\end{eqnarray}
where $s_i$ and $s_f$ represent the initial and final
coordinates in phase space, respectively.
Note that with these simple equations of motion for this Boltzmann equation,
the trajectories are lines, ${\bf p}(s) = {\bf p}_i = {\bf p}_f$ and
\begin{equation}
{\bf r}(s) = {\bf r}_f - {\bf v}(s_f -s) .
\label{eq:path}
\end{equation}
When the starting point of integration is taken to infinity,
$s_i\rightarrow -\infty$, Eq.\ (\ref{eq:gensol}) simplifies to
\begin{equation}
    \tilde g(s_f) = \int_{-\infty}^{s_f}\exp\left( 
    -\int _{s}^{s_f} \frac 1{\tau (s')}ds'
    \right) e\frac{dV}{ds}ds.
    \label{eq:gensol2}
\end{equation}
Equation (\ref{eq:gensol2}) is the starting point for the
calculations in this paper.  Solving it is not always
straightforward because the electro-chemical potential is not
known a priori. In some cases one can deduce $V$ from general
principles.  In other cases Eq.\ (\ref{eq:gensol2}) must be solved
self-consistently as described below.

Once one has a solution to
$\tilde g = \delta g({\bf r}(s),{\bf p}(s))$, the current density, $j$,
can be obtained directly from
\begin{equation}
j_\alpha ({\bf r}) = -e N(E_F) \int \frac {d\theta _p}{2\pi} \, v_\alpha \,
\delta g({\bf r},\theta _p)
\label{eq:current}
\end{equation}
because $\overline g$ has no angular dependence. In this equation
the density of states at the Fermi surface is $N(E_F) = m/(\pi
\hbar ^2)$. For the nonuniform samples considered in this paper,
the current density and the electric field will vary with
position.  An effective conductivity, $\sigma$, for the entire
sample is defined using the spatial average of the current
density, $\langle {\bf j}\rangle$, and the electric field,
$\langle {\bf E}\rangle$,
\begin{equation}
\langle j_\alpha \rangle =
\sigma _{\alpha ,\beta }
\langle E_\beta \rangle .
\label{eq:conductivity}
\end{equation}
This is the conductivity one would obtain by measuring the net
current and voltage drop for a large sample and multiplying by the
usual factors of length and cross-sectional area. In Eq.\
(\ref{eq:conductivity}) the conductivity is defined in terms of
the electric field instead of the gradient of the electro-chemical
potential.  However, because of the periodicity in the
multilayers, $\overline{g}$ is periodic. Thus $\langle \nabla
\overline{g} \rangle=0$ because the spatial average of the
derivative of a periodic function is zero. Consequently, from Eq.\
(\ref{eq:V}) the average electric field, $\langle {\bf E}\rangle$,
is the same as the average gradient in the electro-chemical
potential, $\langle -{\bf \nabla}V\rangle$.

\section{Flat interfaces}

In this section we apply the formalism developed in the previous
section to solve the case of flat interfaces ($A=0$ in Fig.\
\ref{fig:parameters}).  This will provide the basis for the
calculations with wavy interfaces in the next section, since in
this case we are able to solve the Boltzmann equation exactly for
arbitrary mean free paths.  It will also allow us to demonstrate
that this formalism reproduces the conventional results in the
limit of weak spin-flip scattering.

\subsection{CPP geometry}

In the Current-Perpendicular-to-Plane (CPP) geometry the current
flows in the $y$-direction of Figure \ref{fig:parameters}. Current
conservation and translational symmetry in the $x$-direction imply
that the current density is uniform throughout the sample.
Translational symmetry also implies that the electro-chemical
potential, $V$, only depends on the $y$ variable.  The actual
functional form of $V(y)$, however, is not known a priori.

In order to determine $V(y)$ or equivalently ${dV}/{dy}$, the
condition that the spherical average of $\delta g$ is zero (see
Eq.\ (\ref{eq:deltag})) was used in conjunction with Eq.\
(\ref{eq:gensol2}).  This leads to an integral equation for
${dV}/{dy}$.  Because of the periodicity of the multilayers and
hence $dV/dy$, the infinite integral can be converted to a finite
integral.  Discretizing $dV/dy$ then leads to a linear equation,
which is easily solved numerically.

The results of this numerical calculation show that for all mean
free paths the derivative of the electro-chemical potential in the
$i^{\underline{th}}$ layer is proportional to the inverse of $\tau _i$.
Substituting this result into Eq.\ (\ref{eq:gensol2}), it follows
that the distribution function is independent of position and
proportional to ${\bf \hat y}\cdot{\bf \hat p}$, where ${\bf \hat
p}$ is the direction of the momentum.  The current is uniform as
expected. Current uniformity and the fact that $dV/dy$ within a
layer is proportional to $1/\tau _i$ imply that the resistance is
the same as one would obtain by adding classical resistors in series.
The series resistor model for the CPP geometry is commonly used to
analyze experiments in the limit of weak spin-flip scattering.
\cite{TwoCurrentModelBass}

\subsection{CIP Geometry}

In the Current-in-Plane (CIP) geometry the current flows in the
$x$-direction of Figure \ref{fig:parameters}. In this case the
electric field must be constant because of the symmetry of the
problem and ${\bf \nabla \times E}=0$. Furthermore, if we make the
ansatz that $\overline g=0$, then Eq.\ (\ref{eq:gensol2}) can be
integrated analytically. The spherical average of the resulting
distribution function, $\delta g$, is indeed zero, which is
consistent with the ansatz that $\overline g=0$. Thus, in this
case we also have a complete analytic solution for the
distribution function.

\begin{figure}
  {\resizebox{8cm}{!}{\includegraphics{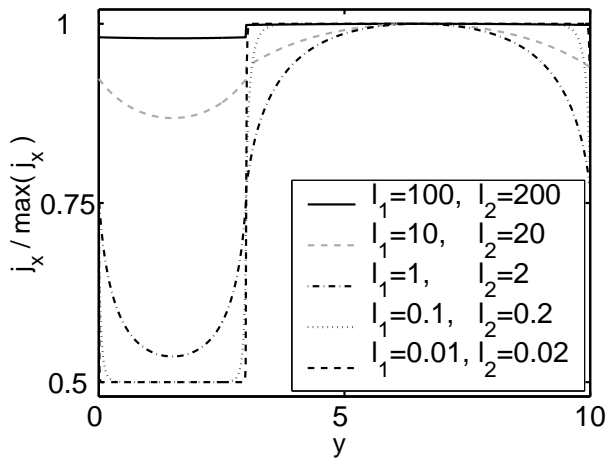}}}
  \caption{\label{fig:CIPcurrent} CIP current density for different
mean free paths in a two-layer repeat
  unit with $\Delta y_1=3$ and $\Delta y_2=7$.
The mean free path in layer $i$ is $l_i = v_F\tau _i$.
For comparison, the current density
  $\mathbf{j}_x$
  for each curve is normalized by
its maximum value,
max($\mathbf{j}_x$).  In the short mean free path
  limit ($\ell_1,\ell_2\ll\Delta y$) the current density changes sharply in
  proportion to $\ell_1/\ell_2$
  across the boundary.  Conversely, in the long mean free path limit
  ($\ell_1,\ell_2\gg\Delta y$) the current density is roughly uniform.  }
\end{figure}

Figure \ref{fig:CIPcurrent} shows the CIP current density in a
two-layer repeat unit multilayer
with $\Delta y_1=3$ and $\Delta y_2 = 7$.
Each curve corresponds to different mean free paths, ranging from
much less than to much greater than $\Delta y$. For comparison,
the curves are normalized by the maximum of $\mathbf{j}_x$. As the
figure illustrates, when the mean free paths are much less than
$\Delta y$, $\mathbf{j}_x$ changes rapidly at the
interface. Moreover, in this limit the ratio of the current
density in layer 2 to that in layer 1 equals the ratio of the
corresponding mean free paths, which is what one would expect
for classical macroscopic resistors in parallel.
As the mean free paths increase,
the current density smooths out, becoming roughly uniform
across the repeat unit as the mean free paths become much greater
than the layer thicknesses.

\subsection{Summary}

In the long mean free path limit the current density and the
distribution functions in both geometries are independent of
position.  As seen in Eq.\ (\ref{eq:gensol2}), there is an average
over a mean free path which goes into computing the distribution
function.  If the mean free path is large enough, then the average
effectively smears out the variations in the sample, producing a
homogeneous current and distribution function.

In the short mean free path limit the current density is the same
as one would obtain classically from adding resistors in series
(CPP geometry) or in parallel (CIP geometry).  In this case the
integral used to calculate the distribution function, Eq.\
(\ref{eq:gensol2}), is short ranged and in particular much smaller
than the layer thicknesses.  Each layer behaves like a macroscopic
piece of metal.

\section{Curved interfaces}

For the case of the wavy interfaces shown in Fig.\
\ref{fig:parameters} it is not possible to obtain simple
analytical expressions for the distribution function like the ones
for the flat interface cases of the previous section. Moreover,
direct numerical solution of the Boltzmann equation requires the
solution of a three-dimensional problem (two space coordinates and
one angular variable along the Fermi surface).  However, we saw in
the previous section that there are two natural limiting cases:
(i) the long mean free path case, where the current density and
distribution function are uniform, and (ii) the short mean free
path case, where the current density and distribution function are
determined by the local gradient in the electro-chemical potential.
In this section we solve for the distribution function, current
density, and conductivity in these two limiting cases.

\subsection{Long mean free path limit}

In the case where the mean free path is long compared to the layer
thicknesses the integral used in computing the distribution
function (Eq.\ \ref{eq:gensol2}) averages the gradient in the
electro-chemical potential over a large region of the sample.  If
the mean free path is long enough, then the gradient in the
electro-chemical potential may be approximated by its average
value. The distribution function then becomes
\begin{eqnarray}
{\tilde g} (s_f) &=& -e {\bf v}\cdot \langle {\bf E}\rangle {\tilde g}'(s_f)
\label{eq:longmfpg}\\
{\tilde g}'(s_f) &=& \int _{-\infty}^{s_f} \exp \left(
-\int _s^{s_f} \frac 1{\tau (s')}ds' \right) ds ,
\label{eq:longmfpgp}
\end{eqnarray}
where the angular brackets denote a spatial average. As discussed
in Sect.\ II, the average of the gradient in the electro-chemical
potential, $\langle -{\bf \nabla} V\rangle$, is equal to the
average electric field, $\langle {\bf E}\rangle$, for a periodic
system.

In the previous section, where we considered flat interfaces, we
did not include interface scattering in the interest of
simplifying the calculation. The scattering rates changed as
electrons went from one material to another, but there was no
additional scattering due to the interfaces.  Here we include
interface scattering since it is crucial for comparing to
experiments.  We model a simplified surface scattering as
infinitesimal layers of a higher-resistivity material.
Surface scattering is included on the right-hand side
of the Boltzmann equation of Eq.\ (\ref{eq:Boltz})
in addition to the bulk scattering term, $1/\tau _b$,
\begin{equation}
\frac 1{\tau({\bf r})} =
\frac 1{\tau _b({\bf r})} +
\Gamma \, \sum _i \int dl 
\delta ^2 ({\bf r} - {\bf R}_i(l)),
\label{eq:surfscattering}
\end{equation}
where the integration runs along the $i^{\underline{th}}$ interface, 
${\bf R}_i(l)$ is the position of the $i^{\underline{th}}$ boundary, 
and $\Gamma$ is a parameter characterizing the
surface scattering rate.
Letting
$\zeta$ be the angle between the velocity, ${\bf v}$, and the unit
vector normal to the boundary with the convention that $0 \le
\zeta \le \pi/2$, the integral in
Eq.\ (\ref{eq:longmfpgp}) from slightly below ($s = t_-$) to
slightly above ($s = t_+$) the interface is
\begin{equation}
\int _{t_-}^{t_+} \frac 1{\tau (s)} ds =
\frac \Gamma {v_F\cos (\zeta )} .
\label{eq:surfscattering2}
\end{equation}
Physically, this means that the probability of an electron being
scattered at an interface is lowest when its trajectory is
perpendicular to the boundary ($\zeta=0$) and larger when it
crosses the boundary at an angle. 

\begin{figure}
  {\resizebox{8cm}{!}{\includegraphics{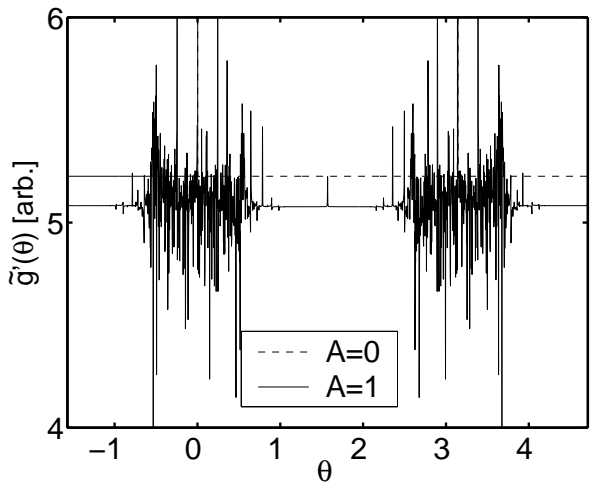}}}
  \caption{\label{fig:gtilde} The function
  $\tilde g^{\prime}(\theta)$
used to compute the distribution function in the long mean free
path limit. The fluctuations present near $\theta = 0$ and $\theta
= \pi$ are due to paths which do not intersect many interfaces.
Once these rapid oscillations are averaged locally, however, this
function is roughly independent of angle and position as expected
in the long mean free path limit. The decrease in $\tilde g'$ as
one goes from flat interfaces ($A=0$) to curved interfaces ($A=1$)
leads to a decrease in the interface conductivity with roughness
in both the CPP and CIP cases. The data shown correspond to a
two-layer repeat unit with $\xi=10$, $\Delta y_1=\Delta y_2=2.5$,
$\ell_1=1000$, $\ell_2=2000$, and $v_F/\Gamma=500$. }
\end{figure}

To test the approximation of Eqs.\ (\ref{eq:longmfpg}) and
(\ref{eq:longmfpgp}) we now evaluate the integral of Eq.\
(\ref{eq:longmfpgp}) numerically. The resulting function, ${\tilde
g}'$, should be independent of both angle and position. Figure
\ref{fig:gtilde} shows $\tilde g^{\prime}(\theta)$ versus
$\theta$ 
for a multilayer composed of two-layer repeat units with $\xi=10$,
$\Delta y_1=\Delta y_2=2.5$, $\ell_1=1000$, $\ell_2=2000$, and
$v_F/\Gamma=500$. The dashed and solid lines represent this sample at
$A=0$ and $A=1$, respectively. The statistical noise present in
the latter curve near $\theta=0$ and $\theta = \pi$ corresponds to
electrons whose trajectories run nearly parallel to the $x$ axis.
Although these electrons intersect significantly fewer interfaces
than those that travel along the $y$ axis, the interfaces can
nevertheless be much more effective in scattering them if they
intersect a boundary nearly tangentially within a few mean free
paths. Therefore $\tilde g^{\prime}$ is highly sensitive to small
changes in $\theta$ near $\theta=0$ and $\theta = \pi$. Despite
the magnitude and frequency of the fluctuations in this region,
the noise averages to the smooth value of $\tilde g^{\prime}$ seen
in the vicinity of $\theta=\pm\pi/2$. This kind of averaging is
performed to compute the current, Eq.\ (\ref{eq:current}), and the
conductivity, Eq.\ (\ref{eq:conductivity}). Thus, once the small
angle fluctuations are averaged the function ${\tilde g}'$ is
independent of angle. We have also verified that it is independent
of position for a given sample.

From Fig.\ \ref{fig:gtilde} it is apparent that ${\tilde g}'$
decreases with increasing interface roughness ($A$), resulting in
a decreasing conductivity in both the CIP and CPP geometries.
Performing the same calculation except without surface scattering
($\Gamma = 0$), one finds that the two values for ${\tilde g}'$
are the same.  In other words, with only bulk scattering in our
model interface roughness does not effect the conductivity.  In
the long mean free path limit, it is therefore necessary to have
interface scattering in order to have a change in the conductivity
with roughness.

In order to examine the additional contribution to the resistance
due to interface scattering, we define an interface conductivity,
$\sigma ^*$, which is extracted as follows.
For a given geometry the conductivities with and without
interface scattering are computed,
$\sigma_{\Gamma\neq 0}$ and $\sigma_{\Gamma=0}$, respectively.
Treating the bulk and interface resistances as resistors in series,
the interface conductivity is given by
\begin{equation}
    \sigma^*=\frac{M}{\Delta y}\bigg{(}\frac{1}{\sigma_{\Gamma\neq0}}
    -\frac{1}{\sigma_{\Gamma=0}}\bigg{)}^{-1},
    \label{eq:sigma*}
\end{equation}
where $M$ is the number of layers in a repeat unit and $\Delta y$
is the thickness of a repeat unit.
In the example above the number of layers is $M=2$ and the length
of a repeat unit is $\Delta y_1 +\Delta y_2 = 5$.
Using a resistors in series model is natural
in the CPP geometry;
however, it may be less clear that this is a good model for the CIP
geometry.  Here, because ${\tilde g}'$ is independent of angle, the changes
in the CPP and CIP conductivities are equal.
Thus, what works in one geometry will also work in the other.

\begin{figure*}
  \subfigure[]{\resizebox{8cm}{!}{\includegraphics{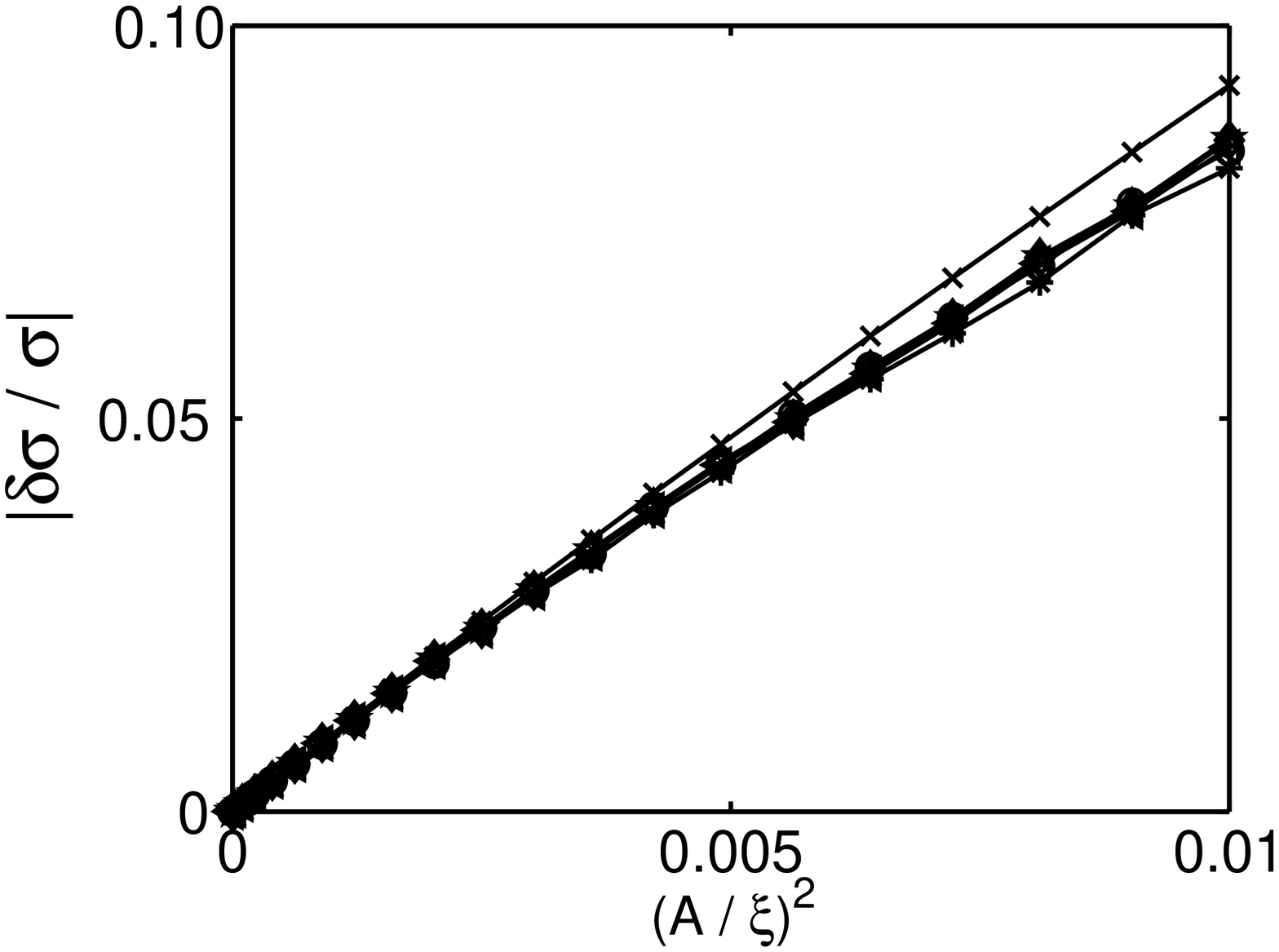}}}
\qquad
  \subfigure[]{\resizebox{8cm}{!}{\includegraphics{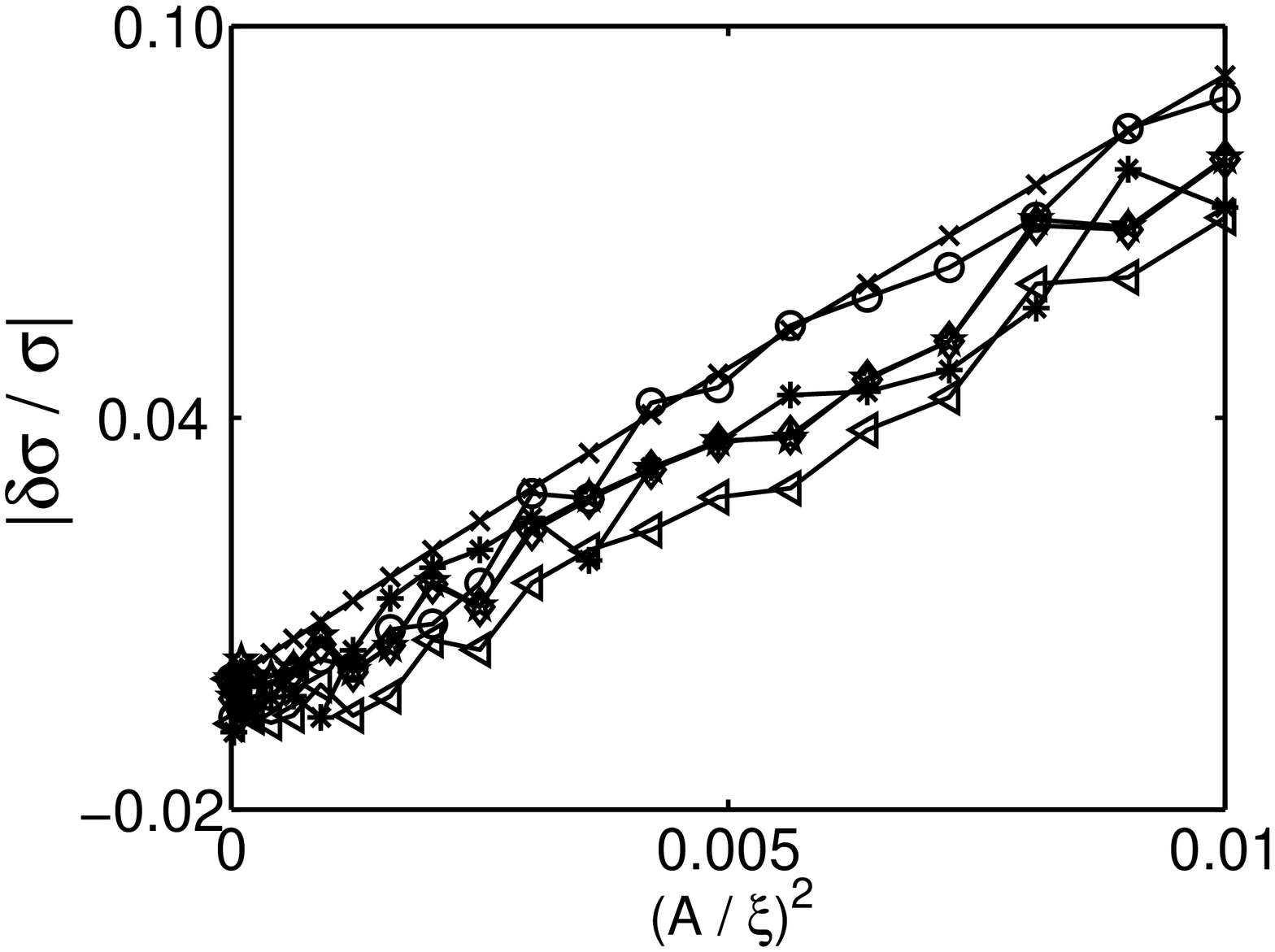}}}
  \caption{\label{fig:percchangeCPP}
  Fractional change in the (a) CPP and (b) CIP interface
  conductivities with roughness in the long mean free path limit.
  Also shown is the corresponding change in interface
  length ($\times$) with roughness.
  In both cases the fractional change in the interface conductivity,
  $|\delta \sigma ^*/\sigma ^*|$, is very close to the fractional increase
  in the length of the interface,
  ${\delta \mathcal{L}}/{\mathcal{L}}$.
  In our model, which assumes sinusoidal interfaces, both of these
  quantities are proportional to $(A/\xi )^2$. The ($\ast$),
  ($\circ$), and ($\diamond$) data points represent
  different coordinates within a sample that has $\Delta y_1=\Delta
  y_2=2.5$, $\ell_1=1000$, $\ell_2=2000$, and $v_F/\Gamma=500$.  
  The ($\star$) and ($\triangleleft$) points
  represent the same coordinate as the ($\diamond$) points, but correspond to
  samples with $\ell_1=\ell_2=1000$
  and twice the surface scattering rate, respectively.
  The larger scatter in the data
  for the CIP case is due to the fluctuations in $\tilde g^{\prime}$
  in Fig.\ \ref{fig:gtilde}. }
\end{figure*}

For a given set of bulk and surface scattering rates define
$\delta \sigma ^*$ as the difference between the interface
conductivity for $A \ne 0$ and $A = 0$,
\begin{equation}
\delta \sigma ^* = \sigma ^* (A) - \sigma ^*(0).
\label{eq:deltasigma*}
\end{equation}
Figures \ref{fig:percchangeCPP} (a) and (b) contain plots of
$|\delta \sigma^*/\sigma^*|$ versus $(A/\xi)^2$ for the CPP and
CIP geometries, respectively. From Eq.\ (\ref{eq:longmfpg}) one
can see that the distribution function $\delta g = {\tilde g}$ in
the CIP case, where the average electric field is in the
$x$-direction, is largest near $\theta = 0$ and $\theta = \pi$,
while in the CPP case, $\delta g$ is largest near $\theta =
\pm\pi/2$.  The larger scatter in the data of Fig.\
\ref{fig:percchangeCPP} for the CIP case than the CPP case
reflects the larger scatter in $\tilde g'$ near $\theta = 0$ and
$\theta = \pi$.

The curves in these figures represent
either various positions within a particular sample or fixed positions
within samples that
have different mean free paths or surface scattering rates.
When plotted in this manner, all of the data fall onto the same line.
Thus, not only is $\delta \sigma ^*/\sigma ^*$ proportional to
$(A/\xi)^2$, but the proportionality constant is independent of
the model parameters.

One difference between a curved interface and a flat interface is
that the curved one is longer.  Because of surface scattering, a
curved interface will have more scattering and hence a larger
resistance. The amount of extra scattering provided by a curved
interface should be proportional to the additional length in the 
boundary. Let $\mathcal{L}$ be the length of the
interface and $\delta \mathcal{L}$ be the change in the interface
length from the flat case ($A=0$) to the wavy case ($A\ne 0$).  The
percent increase in the interface length, $\delta
\mathcal{L}/\mathcal{L}$, is plotted as the $\times$'s in Fig.\
\ref{fig:percchangeCPP}. To a good approximation it is evident
from Fig.\ \ref{fig:percchangeCPP} that the percent decrease in
the interface conductivity is equal to the percent increase in the
length of the interface. For our sinusoidal boundaries the
percentage change in the interface length is $\pi ^2 (A/\xi )^2$
for small $A$, so within our model we have
\begin{equation}
    \frac{\delta \sigma^*}{\sigma^*}\approx
-\frac{\delta \mathcal{L}}{\mathcal{L}}
    \approx-\pi^2\bigg{(}\frac{A}{\xi}\bigg{)}^2.
    \label{eq:percchange}
\end{equation}

\subsection{Short mean free path limit}

In the short mean free path limit the integral used to compute the
distribution function, Eq.\ (\ref{eq:gensol2}), samples the
gradient in the electro-chemical potential over a short distance.
Thus, we may approximate the electro-chemical potential as constant
within the integral, and the distribution function becomes
\begin{equation}
    \delta g({\bf r},{\bf\hat p}) \approx e \tau v_F {\bf \hat p} \cdot
    {\bf \nabla}V({\bf r}) .
    \label{eq:gensolwavyshort}
\end{equation}
In this approximation, the current density and conductivity in the $i$-th layer
are given by
\begin{eqnarray}
{\bf j}({\bf r}) &=& \sigma ({\bf r}) (-{\bf \nabla}V({\bf r})),
\label{eq:currentshortmfp}\\
\sigma ({\bf r}) &=& \frac {ne^2\tau (\bf r)}m .
\label{eq:localconductivity}
\end{eqnarray}
Since the conductivity is constant within each layer,
the current conservation condition, ${\bf \nabla \cdot j} = 0$,
within a given layer implies that
$ -\nabla^2 V_i=0$,
where $V_i$ is the electro-chemical potential in layer $i$.
The problem
of finding the current density and effective conductivity due to
roughness thus reduces to solving Laplace's equation within each
layer subject to the boundary conditions that the current is conserved and
that the electro-chemical potential is continuous.
As noted earlier,
the average electric field, $\langle {\bf E} \rangle$, is the
same as the average of the gradient in the electro-chemical potential,
$-{\bf \nabla}V$, because of the periodicity of the multilayers.

Whereas in the long mean free path limit we had to include interface
scattering or else there would not have been any effect, in the
short mean free path limit there is already an effect without surface
scattering.  From calculations which include both bulk and surface
scattering, we have found that the presence of interface scattering does
not qualitatively alter the results for the conductivity.
Thus, in the results presented below we do not include interface scattering.

\begin{figure}
  {\resizebox{8cm}{!}{\includegraphics{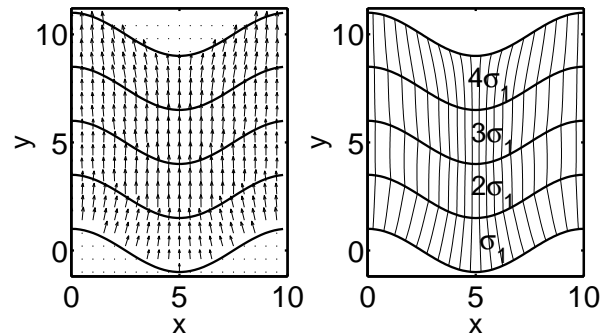}}}
  \caption{\label{fig:CPPcurrent} CPP current density vector field and
  current field lines in the short mean free path limit.  Due to roughness,
current tends
  to traverse a greater distance in high conductivity than low conductivity
  layers, leading to an increase in the effective CPP conductivity with
roughness.  }
\end{figure}

Figure \ref{fig:CPPcurrent} shows the CPP current density vector field
and current field lines in a repeat unit with $A=1$, $\Delta
y=\xi=10$, $\Delta y_i=2.5$, and $\sigma_i=i\sigma_1$.  As the
figure illustrates, roughness causes current to flow nonlinearly
through the sample in such a way that it traverses a greater
distance through the high conductivity layers than through the low
conductivity layers. The effect of roughness is thus to increase
the effective CPP conductivity, $\sigma $, of Eq.\
(\ref{eq:conductivity}).

\begin{figure}
  {\resizebox{8cm}{!}{\includegraphics{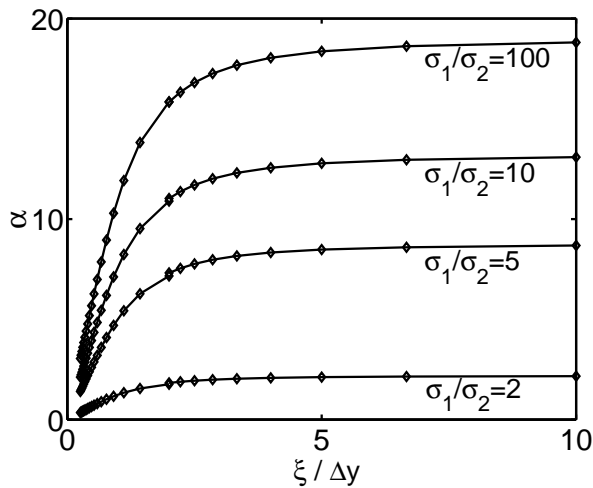}}}
  \caption{\label{fig:alpha} The proportionality constant $\alpha_{CPP}$
between $\delta \sigma /\sigma $ and $(A/\xi
)^2$ in the CPP geometry.  This constant $\alpha _{CPP}$ increases
  with the ratio of the layer conductivities
and $\xi/\Delta y$, saturating for $\sigma_1\gg\sigma_2$
  and when $\xi$ becomes several times $\Delta y$.  }
\end{figure}

As in the long mean free path limit, the percentage increase in
$\sigma$ relative to the flat interface case is proportional to
$(A/\xi)^2$,
\begin{equation}
\frac {\delta \sigma}{\sigma} \approx \alpha \left( \frac A\xi
\right)^2. \label{eq:alphadef}
\end{equation}
Here, however, the proportionality constant, which we denote as
$\alpha_{CPP}$ for the CPP case,
depends on both the geometry and the layer
conductivities. Figure \ref{fig:alpha} shows $\alpha_{CPP}$ as a
function of $\xi/\Delta y$ for several values of
$\sigma_1/\sigma_2$ for a sample composed of two-layer repeat units
with $\Delta y_1 = \Delta y /2$.
As the figure illustrates, $\alpha_{CPP}$
increases with the ratio of the layer conductivities and saturates
for $\sigma_1\gg\sigma_2$.  It also increases
with $\xi/\Delta y$, saturating as $\xi$ becomes several times
$\Delta y$.

\begin{figure}
  {\resizebox{8cm}{!}{\includegraphics{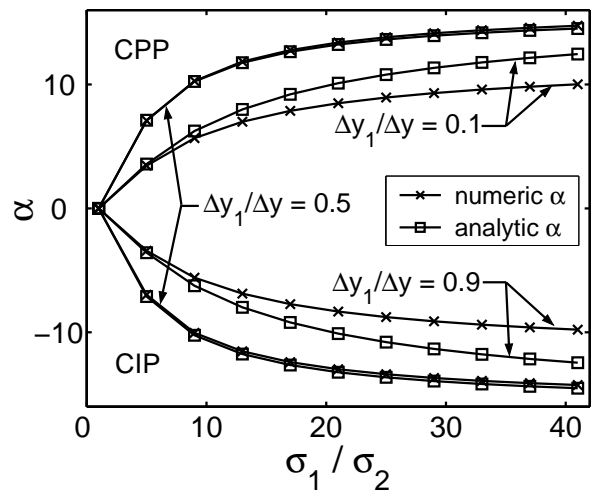}}}
  \caption{\label{fig:alphaLevy}
Comparison of $\alpha$ for the CPP and CIP geometries.
  The numeric CPP and CIP curves shown are roughly mirror 
images of one another,
illustrating that
  $\alpha_{CIP}(\sigma_1,\sigma_2)\approx-\alpha_{CPP}(\sigma_2,\sigma_1)$.
Depending on the sample geometry, the analytical approximations
for $\alpha$ in Eqs.\ (\ref{eq:alphaCIP}) and (\ref{eq:alphaCPP})
are excellent in some cases and only roughly correct in others.
The data shown are for a two layer repeat unit with $\xi = 10$ and
$\Delta y = 5.2$.}
\end{figure}

\begin{figure}
  {\resizebox{8cm}{!}{\includegraphics{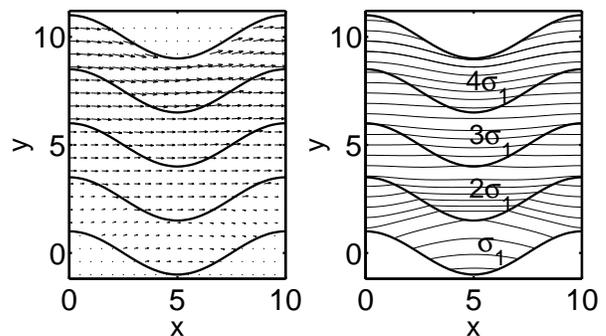}}}
  \caption{\label{fig:CIPcurrent2} CIP current density vector field and
  current field lines in the short mean free path limit.  As a result of
  roughness, current near an interface does not remain within a single layer
  across a period but traverses both high and low conductivity regions.
This reduces the short circuit effect of the high conductivity
layers and leads to a decrease in the conductivity of the
multilayer. }
\end{figure}

To determine the current and conductivity in the CIP geometry,
the same Laplace's equation is solved with the net potential
drop in the $x$ as opposed to the $y$ direction.
Figure \ref{fig:CIPcurrent2} shows the CIP current density vector
field and current field lines in a four-layer repeat unit with
$A=1$, $\Delta y=\xi=10$, $\Delta y_i=2.5$, and
$\sigma_i=i\sigma_1$.  Due to roughness, electrons in high
conductivity layers impinge on lower conductivity layers near the
interfaces.
This in effect reduces the short circuit effect of the high conductivity
layers and consequently decreases the effective CIP conductivity.

The percent change in the effective conductivity relative to the
flat case ($A=0$) is again proportional to $(A/\xi)^2$. The
proportionality constant, $\alpha_{CIP}$, is simply related to
$\alpha_{CPP}$ in the two-layer case: $\alpha_{CIP}$ approximately
equals the negative of $\alpha_{CPP}$, where the latter is
calculated with the layer conductivities interchanged
($\alpha_{CIP}(\sigma_1,\sigma_2)\approx-\alpha_{CPP}(\sigma_2,\sigma_1)$).
Equivalently, $\alpha_{CIP}$ is roughly equal to the negative of
$\alpha_{CPP}$ when the latter is computed for a sample in which
the thicknesses of layers 1 and 2 are interchanged
($\alpha_{CIP}(\Delta y _1,\Delta y _2)\approx-\alpha_{CPP}(\Delta
y _2,\Delta y_1)$). This result is illustrated in Fig.\
\ref{fig:alphaLevy}, where the numeric $\alpha_{CPP}$ versus
$\sigma_1/\sigma_2$ curves for $\Delta y_1/\Delta y=0.5$ and $0.1$
are mirror images of the numeric $\alpha_{CIP}$ versus
$\sigma_1/\sigma_2$ curves for $\Delta y_1/\Delta y=0.5$ and
$0.9$.

Approximate analytic expressions for both $\alpha _{CIP}$ and
$\alpha _{CPP}$ can be derived using the approach of Levy \emph{et
al}.\ for grooved multilayers \cite{Levy}.  These multilayers are
similar to those shown in Fig.\ \ref{fig:parameters} except that
the interfaces are piecewise linear.  The angle the interfaces
make with the $x$-axis is defined to be $\theta$, and it is
assumed that the layers are far apart relative to the amplitude of
the interface fluctuations, which we have called $A$ in the
sinusoidal case. In the CIP configuration they approximate the
electric field to be uniform in the $x$-direction. Letting
$\sigma_{CIP}$ and $\sigma_{CPP}$ be the CIP and CPP
conductivities when the interfaces are flat, it follows that the
spatial average of the current density is
\begin{equation}
\langle j_x \rangle = \left (
   \sigma _{CIP} \cos ^2\theta +
   \sigma _{CPP} \sin ^2\theta \right) E_x.
\label{eq:jCIP}
\end{equation}
Consequently, the effective conductivity is
\begin{equation}
\sigma = \sigma _{CIP}\cos ^2\theta + \sigma _{CPP}\sin ^2 \theta.
\label{eq:sigmaCIP}
\end{equation}
Although we consider sinusoidal as opposed to piecewise linear
interfaces, we can use an approximate effective $\theta $ of
$2A/(\xi /2) = 4A/\xi$. Expanding Eq.\ (\ref{eq:sigmaCIP}) for
small $\theta$ and using the definition in Eq.\
(\ref{eq:alphadef}), it follows that $\alpha$ in the CIP geometry
is
\begin{equation}
\alpha _{CIP} \approx 16 \left( \frac{\sigma _{CPP}}{\sigma _{CIP}} -1\right) .
\label{eq:alphaCIP}
\end{equation}

Figure \ref{fig:alphaLevy} shows $\alpha_{CIP}$ versus
$\sigma_1/\sigma_2$ for different values of $\Delta y_1/\Delta y$
as computed analytically via Eq.\ (\ref{eq:alphaCIP}) and
numerically in the short mean free path limit. The data correspond
to a sample composed of two-layer repeat units with $\xi=10$ and
$\Delta y=5.2$.  As the figure illustrates, the agreement between
the analytic and numeric $\alpha_{CIP}$ is excellent for some
parameters and worse for others. In the case of two-layer repeat 
units, we find the agreement
to be particularly good when the geometric parameters satisfy
$\xi/\Delta y_1 \approx 4$ for $\sigma_1\geq\sigma_2$ and 
$\xi/\Delta y_2 \approx 4$ for $\sigma_1\leq\sigma_2$. This relation is
satisfied by the CIP curves corresponding to $\Delta y_1/\Delta
y=0.5$ in Fig.\ \ref{fig:alphaLevy}.  In any case, the analytic
expression does provide an estimate for $\alpha_{CIP}$.

In the CPP case if one starts from Eq.\ (\ref{eq:sigmaCIP}) to
obtain $\alpha _{CPP}$, one obtains a poor fit to the numerical
solution. In the CIP geometry for the case of flat interfaces,
symmetry required that the electric field be uniform.  On the
other hand, in the CPP geometry for the case of flat interfaces,
symmetry required that the current density be uniform. Thus, in
the CPP geometry we start from the assumption that the current
density is uniform in the $y$-direction.  The average electric
field in the $y$-direction is then
\begin{equation}
\langle E_y \rangle = \left (
   \rho _{CPP} \cos ^2\theta +
   \rho _{CIP} \sin ^2\theta \right) j_y,
\label{eq:ECPP}
\end{equation}
where $\rho_{CPP}$ and $\rho_{CIP}$ are the CPP and CIP
resistivities corresponding to flat interfaces. The effective
resistivity is therefore
\begin{equation}
\rho = \rho _{CPP}\cos ^2\theta + \rho _{CIP}\sin ^2 \theta
\label{eq:rhoCPP}
\end{equation}
Note that this equation is not the same as Eq.\ (\ref{eq:sigmaCIP})
from Ref. \cite{Levy}, but rather its generalization to
the case of the CPP geometry.  
As will be shown below, it provides a much better to fit to our
numerical data on the CPP resistivity.
Equation (\ref{eq:rhoCPP})
and the assumption that $\theta \approx 4A/\xi$
imply that the CPP proportionality
constant is the negative of the CIP proportionality constant,
\begin{equation}
\alpha _{CPP} = -\alpha _{CIP},
\label{eq:alphaCPP}
\end{equation}
independent of the sample geometry.  Note, however, that
numerically Eq.\ (\ref{eq:alphaCPP}) holds for two-layer repeat
units only when $\Delta y_1=\Delta y_2$.

The analytic expression for $\alpha _{CPP}$ is compared to the
numerical solution in Fig.\ \ref{fig:alphaLevy}. As for the CIP
case, the analytic estimates are excellent for certain geometries
but not as good for others.  We find the agreement 
in the case of two-layer repeat units to be
particularly good when the geometric parameters satisfy
$\xi/\Delta y_2\approx 3.7$ for $\sigma_1\geq\sigma_2$ and 
$\xi/\Delta y_1\approx 3.7$ for $\sigma_1\leq\sigma_2$ . The CPP curves
corresponding to $\Delta y_1/\Delta y=0.5$ in Fig.\
\ref{fig:alphaLevy} satisfy this relation.

\subsection{Comparison of the long and short mean free path limits}

The physics of the change in the effective conductivity due to
long length scale interface roughness is different in the long and
short mean free path limits. The effect in the long mean free path
limit is due entirely to enhanced interface scattering with
roughness, and no effect is observed when surface scattering is
ignored. In the short mean free path limit, interface scattering
plays a less dominant role, as the effective conductivity changes with
roughness even when only bulk scattering is present. Here the
effect in the CPP geometry results from roughness providing a
less-resistive, non-linear path of current flow, while the effect
in the CIP geometry is due to a reduction of the short circuit
effect of the high conductivity layers.

In the long mean free path limit, the decrease in the interface
conductivity due to roughness approximately equals the
corresponding increase in the interface length in both the CPP and
CIP geometries. Thus, in our model we have
\begin{equation}
    \frac{\delta \sigma^*}{\sigma^*} \approx
    -\frac{\delta\mathcal{L}}{\mathcal{L}}
    \approx-\pi^2\bigg{(}\frac{A}{\xi}\bigg{)}^2.
    \label{eq:longmfp}
\end{equation}
In the short mean free path limit, the percent change in the
effective conductivity due to roughness is also proportional to
$(A/\xi)^2$,
\begin{equation}
    \frac{\delta \sigma}{\sigma}\approx
    \alpha\bigg{(}\frac{A}{\xi}\bigg{)}^2.
    \label{eq:shortmfp}
\end{equation}
Here, however, the proportionality constant $\alpha$ depends on
the layer conductivities and the geometry of the sample.
For a multilayer composed of two-layer repeat units with $\xi=10$
and $\Delta y_1=\Delta y_2=2$, the coefficient $|\alpha |$ is
approximately 2 for $\sigma _1/\sigma _2 =2$, 12 for $\sigma
_1/\sigma _2 = 10$), and 18 for $\sigma _1 \gg \sigma _2$,
indicating a range of order 10.

Within each limit, we make predictions concerning distinct
physical quantities -- interface conductivity versus effective
conductivity. To compare the magnitude of the percentage change in
the effective conductivity in both limits, we write $\delta
\sigma/\sigma$ in the long mean free path limit in terms of
$\delta \sigma^*/\sigma^*$, where
\begin{equation}
  \sigma =\bigg{(}\frac{1}{\sigma_{\Gamma=0}}+\frac{M}{\Delta
  y}\frac 1{\sigma ^*}\bigg{)}^{-1}.
  \label{eq:sigma*longmfp}
\end{equation}
Using Eq.\ (\ref{eq:sigma*longmfp}) and the fact that $\sigma_{\Gamma=0}$
is independent of $A$, we find that
\begin{equation}
    \frac{\delta \sigma}{\sigma}=
    \frac{\delta \sigma^*}{\sigma^*}\bigg{(} 1+ \frac{\Delta y
    \sigma^*}{M \sigma_{\Gamma=0}}\bigg{)}^{-1}.
    \label{eq:longmfp2}
\end{equation}
Since the second term in the parenthesis is greater than or equal
to 0, $\delta \sigma/\sigma$ must be less than or equal to $\delta
\sigma^*/\sigma^*$, where equality holds in the limit of zero bulk
resistance. Additionally, $\delta \sigma/\sigma$ goes to zero in
the absence of surface scattering.  The constant of
proportionality between $\delta \sigma/\sigma$ and $(A/\xi)^2$
therefore lies between 0 and $-\pi^2$ in the long mean free path
limit. Comparing this range to the values of $|\alpha|$ stated
above, we conclude that the magnitude of the percent change in the
effective conductivity due to roughness can be greater in either
the long or short mean free path limits.

\section{Giant Magnetoresistance}

Magnetic multilayers are alternating layers of ferromagnetic and
paramagnetic metals.  For a large field in the plane of the
layers, the magnetic moments of the ferromagnetic layers align,
creating the parallel magnetic configuration (P). At zero or a
lower field, the magnetic moments of the layers are not aligned in
parallel. With a proper choice for the thickness of the
paramagnetic layers, the ferromagnetic layers may be coupled
antiferromagnetically so that adjacent magnetic layers have
moments pointing in opposite directions. We call this the
antiparallel (AP) configuration. The resistance of the multilayer
in the two magnetic configurations is different, leading to a
magnetoresistance. It is common to characterize this
magnetoresistance as a ratio called the giant magnetoresistance or
GMR,
\begin{equation}
    GMR = \frac{\rho^{AP}-\rho^P}{\rho^P},
    \label{eq:GMR}
\end{equation}
where $\rho^{AP}$ and $\rho^{P}$ are the resistivities of the antiparallel
and parallel configurations.

In many systems the spin relaxation length, which is how far an
electron's spin maintains its orientation, is long compared to the
thickness of the layers. Hence, it is often possible to
approximate the resistance as being due to two parallel channels
for conduction, one channel for each of the two possible spin
orientations. In the following we make this approximation, the
corrections to which are well know in both the CPP and CIP
geometries \cite{ValetFert,SpinFlipCIP}.

\subsection{Long Mean Free Path Limit}

In order to deduce how the CPP GMR is effected by roughness in the
long mean free path limit, within each spin channel we treat the
layers as resistors in series. First, let $t_{PM}$ and $t_{FM}$ be
the thicknesses of the paramagnetic and ferromagnetic layers,
respectively, and let $\Delta y=2(t_{PM}+t_{FM})$. The resistivity
for electrons of either spin in the paramagnetic material is
denoted by $\rho_{PM}$, and the resistivity of the majority/minority
electrons in the ferromagnetic material is $\rho_{FM,maj/min}$.
The interface resistance for the majority/minority electrons is
$\rho^*_{maj/min}$. Next, define $\rho_{maj}$ and $\rho_{min}$ by
\begin{eqnarray}
    \Delta y \rho_{maj}&\equiv&
    2t_{PM}\rho_{PM}+2t_{FM}\rho_{FM,maj}+4\rho^*_{maj}
    \label{eq:rhomaj} \\
    \Delta y \rho_{min}&\equiv&
    2t_{PM}\rho_{PM}+2t_{FM}\rho_{FM,min}+4\rho^*_{min}.
    \label{eq:rhomin}
\end{eqnarray}
In the parallel magnetic configuration, the resistivity of the
majority spin channel is $\rho _{maj}$, and the resistivity of the
minority spin channel is $\rho _{min}$.  Adding the resistivity of
the two spin channels in parallel gives a resistivity for the
parallel configuration of
\begin{equation}
    \rho^P_{CPP} = \frac{\rho_{min}\rho_{maj}}{\rho_{min}+\rho_{maj}} .
    \label{eq:rhoP}
\end{equation}
When the layers are aligned antiferromagnetically, both spin
channels have the same resistivity, $(\rho_{maj}+\rho_{min})/2$.
Adding the two spin channels in parallel gives an antiparallel
configuration resistivity of
\begin{equation}
    \rho^{AP}_{CPP}=\frac{1}{4}(\rho_{maj}+\rho_{min}) .
    \label{eq:rhoAP}
\end{equation}

As we have seen in the previous section, the effect of interface
roughness in the long mean free path limit can be expressed as an
increase in the interface resistance, $\delta \rho ^*$. To
determine the effect of interface roughness on the parallel
resistivity, the antiparallel resistivity, and the GMR, we expand
Eqs.\ (\ref{eq:rhoP}), (\ref{eq:rhoAP}), and (\ref{eq:GMR}) to
linear order in the changes in the interface resistances, $\delta
\rho ^*_{maj}$ and $\delta \rho ^*_{min}$:
\begin{eqnarray}
    (t_{FM}+t_{PM})\delta\rho^P &=&
    \frac{2\rho_{min}^2}{(\rho_{maj}+\rho_{min})^2}\delta
    \rho^*_{maj} \nonumber \\
    &+&\frac{2\rho_{maj}^2}{(\rho_{maj}+\rho_{min})^2}\delta
    \rho^*_{min} \label{eq:Pexpansion}\\
    (t_{FM}+t_{PM})\delta\rho^{AP} &=& \frac{1}{2}\delta
    \rho^*_{maj}+\frac{1}{2}\delta \rho^*_{min} \label{eq:APexpansion} \\
   \delta GMR &=&
    \frac{1}{\Delta y}
    \bigg{(}\frac{1}{\rho_{maj}^2}
    -\frac{1}{\rho_{min}^2}\bigg{)} \times \label{eq:GMR2} \\
    & & \times (\rho_{maj}\delta\rho^*_{min}-
    \rho_{min}\delta\rho^*_{maj}). \nonumber
\end{eqnarray}
The changes in the interface resistances due to long length scale
disorder may be obtained from Eq.\ (\ref{eq:longmfp}) using the
fact that $\rho ^*$ is the inverse of $\sigma ^*$ for both the
majority and the minority electrons,
\begin{equation}
    \frac{\delta \rho^*_{maj/min}}{\rho^*_{maj/min}} \approx -
    \frac{\delta \sigma^*_{maj/min}}{\sigma^*_{maj/min}}
    \approx \pi^2\bigg{(}\frac{A}{\xi}\bigg{)}^2.
    \label{eq:longmfp3}
\end{equation}

From Eq.\ (\ref{eq:longmfp3}) we can see that $\delta \rho ^*$ is
positive for both the minority and majority electrons.
Consequently, in the long mean free path limit both $\delta \rho
^P$ and $\delta \rho ^{AP}$ are positive, i.e., they increase with
increasing long length scale interface disorder. On the other
hand, the second term on the right hand side in Eq.\
(\ref{eq:GMR2}) can be either positive or negative depending on
the values of the resistivities and interface resistances. Hence
when the electronic mean free paths are much greater than the
layer thicknesses, the CPP GMR can either be enhanced or reduced
by roughness depending on the sample.  The same holds true for the
CIP GMR since in the long mean free path limit the CPP and CIP
resistances are the same.

\subsection{Short mean free path limit}

To compute the GMR in the short mean free path limit we consider a
four-layer repeat unit with layers 1 and 3 being ferromagnetic and
layers 2 and 4 being paramagnetic.  As in the previous section, in
the short mean free path limit we do not include interface
scattering because it does not change the results qualitatively.
Let $t_{FM}=\Delta y_1=\Delta y_3$ be the thickness of the
ferromagnetic layers and $t_{PM}=\Delta y_2=\Delta y_4$ be the
thickness of the paramagnetic layers. The conductivity for either
spin direction in the paramagnetic layers is defined as $\sigma
_{PM}$, and the conductivities for the majority and minority spins
in the ferromagnetic material are $\sigma _{FM,maj}$ and $\sigma
_{FM,min}$, respectively.

In the parallel magnetic configuration, the majority band
electrons experience conductivities of $\sigma _1 = \sigma _3 =
\sigma _{FM,maj}$ and $\sigma _2 = \sigma _4 = \sigma _{PM}$.
Similarly, the minority band electrons experience conductivities
of $\sigma _1 = \sigma _3 = \sigma _{FM,min}$ and $\sigma _2 =
\sigma _4 = \sigma _{PM}$. When the magnetizations of adjacent
ferromagnetic layers are aligned antiparallel, both spin channels
have the same net conductivity since the ferromagnetic layers (1
and 3) have the same thicknesses and the paramagnetic layers (2
and 4) have the same thicknesses.  The conductivity for either
spin channel can thus be computed using $\sigma _1 = \sigma
_{FM,maj}$, $\sigma _3 = \sigma _{FM,min}$, and $\sigma _2 =
\sigma _4 = \sigma _{PM}$.

Solving Laplace's equation with the boundary conditions described
in Sect.\ IV B, we compute numerically the conductivities
for the parallel and antiparallel configurations, $\sigma _P$ and
$\sigma _{AP}$. The giant magnetoresistance then follows from Eq.\
(\ref{eq:GMR}), which when expressed in terms of the conductivity
becomes $(\sigma _P - \sigma _{AP})/\sigma _{AP}$. 
In contrast to the CPP conductivity in either the parallel or
antiparallel magnetic configurations, which always increases with
long length scale interfacial roughness, we find that the CPP 
GMR can either increase or 
decrease with $(A/\xi)^2$ depending on the geometry of the 
sample and the layer conductivities.  Specifically, for a sample with
a given 
$t_{FM}$, $t_{PM}$, and $\sigma_{FM,maj/min}$, the GMR initially 
increases with roughness as 
$\sigma_{PM}$ is increased from zero.  At some point below
$(\sigma_{FM,maj}+\sigma_{FM,min})/2$, $\sigma_{PM}$ reaches a
critical value at which the GMR is independent of roughness.  
The GMR subsequently 
decreases with roughness as $\sigma_{PM}$ is increased beyond this
value.  
This critical value, $\sigma_{PM}^c$, depends on the layer thicknesses
and $\sigma_{FM,maj/min}$ and has the form 
\begin{equation}
	\sigma_{PM}^c = C(1-\exp(-rt_{PM}/\Delta y)),
\end{equation}
where $C$ is a constant slightly below
$(\sigma_{FM,maj}+\sigma_{FM,min})/2$ and $r$ is a constant of order 10,
The exact value of these constants depends on $\sigma_{FM,maj/min}$
and the layer thicknesses.

\begin{figure} [h!]
  {\resizebox{8cm}{!}{\includegraphics{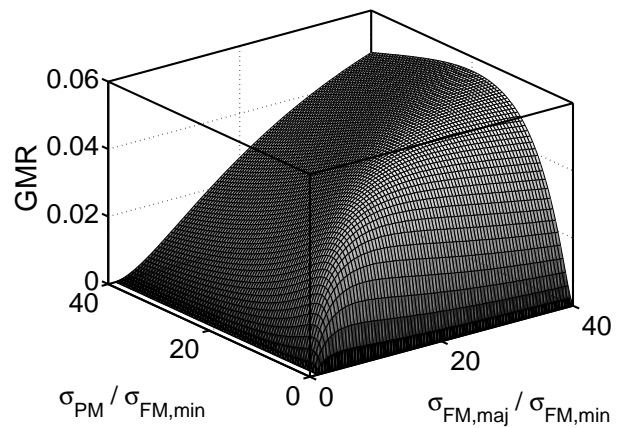}}}
  \caption{\label{fig:GMR} Current-in-Plane GMR in the short
mean free path limit.
  The presence of
  interfacial roughness leads to a giant magnetoresistance which is
proportional to $(A/\xi)^2$ and thus vanishes only for $A = 0$.
The conductivities of the paramagnetic layer (for either spin),
the ferromagnetic majority bands, and the ferromagnetic minority
bands are $\sigma _{PM}$, $\sigma _{FM,maj}$, and $\sigma
_{FM,min}$, respectively. The maximum GMR is achieved when $t_{PM}$ is
  small compared to $t_{FM}$ and for
  $\sigma_{FM,maj}\gg\sigma_{FM,min}$ and
  $\sigma_{PM}\approx\frac{1}{2}\sigma_{FM,maj}$.
The data shown correspond to a multilayer with $A = 1.5$,
$\xi=10$, $t_{FM} = 5.33$, and $t_{PM} = 2.67$. }
\end{figure}

The CIP GMR in the short mean free path limit can be computed in
the same manner as the CPP GMR by just changing the boundary
condition from an applied field in the $y$-direction to an applied
field in the $x$-direction. Here, we find the CIP GMR to be
positive and proportional to $(A/\xi)^2$. In the short mean free
path limit, the CIP GMR therefore vanishes only when the
interfaces are flat ($A=0$). Figure \ref{fig:GMR} shows the CIP
GMR as a function of $\sigma_{PM}/\sigma_{FM,min}$ and
$\sigma_{FM,maj}/\sigma_{FM,min}$ for a multilayer with $A=1.5$,
$\xi=10$, $t_{FM}=5.33$, and $t_{PM} = 2.67$.  The CIP GMR
increases as $t_{PM}$ decreases, and within a particular
geometry the GMR reaches its maximum value when 
$\sigma_{FM,maj}\gg\sigma_{FM,min}$ and 
$\sigma_{PM}\approx\frac{1}{2}\sigma_{FM,maj}$.

In addition to solving for the conductivities exactly in the short
mean free path limit, one can also use the approximate expressions
of Eqs.\ (\ref{eq:alphadef}), (\ref{eq:alphaCIP}), and
(\ref{eq:alphaCPP}). Although not as accurate as the numerical
solution, this technique may be useful in estimating the size of
the effect for an actual experiment.

According to these equations the percent change in the
conductivity due to long length scale surface roughness is
proportional to the ratio of the CPP and CIP conductivities
corresponding to flat interfaces. There are two possible magnetic
orientations of the layers, parallel and antiparallel. For each of
these magnetic orientations the electrons can either be in the
minority or majority bands. Thus, since there are two geometries
(CPP or CIP), two magnetic orientations (P or AP), and two spin
channels (maj or min), there is a total of eight conductivities to
be specified. We denote the conductivity of the majority band for
the CPP geometry in the parallel magnetic orientation as $\sigma
^{P}_{CPP,maj}$ and label the other conductivities in a similar
fashion.

For the CPP geometry the conductivities can be computed by adding
the resistances of the layers in series.  Using the notation for
$\rho _{maj}$ and $\rho _{min}$ in Eqs.\ (\ref{eq:rhomaj}) and
(\ref{eq:rhomin}) with $\rho ^* = 0$, the CPP conductivities are
\begin{eqnarray}
\sigma ^P_{CPP,maj} &=& 1/\rho _{maj} \label{eq:sigmaPCPPmaj}\\
\sigma ^P_{CPP,min} &=& 1/\rho _{min} \label{eq:sigmaPCPPmin}\\
\sigma ^{AP}_{CPP,maj} =
\sigma ^{AP}_{CPP,min} &=& \frac 2{\rho _{min}+\rho _{maj}}
                         \label{eq:sigmaAPCPP} .
\end{eqnarray}
For the CIP geometry the conductivities can be computed by adding
the resistances of the layers in parallel. It is useful to define
the analog of Eqs.\ (\ref{eq:rhomaj}) and (\ref{eq:rhomin}) for
the layers in the parallel case,
\begin{eqnarray}
    \Delta y \sigma_{maj}&\equiv&
    2t_{PM}\frac 1{\rho_{PM}}+2t_{FM}\frac 1{\rho_{FM,maj}}
    \label{eq:sigmamaj} \\
    \Delta y \sigma_{min}&\equiv&
    2t_{PM}\frac 1{\rho_{PM}}+2t_{FM}\frac 1{\rho_{FM,min}}.
    \label{eq:sigmamin}
\end{eqnarray}
Note that with this notation $\rho _{maj} \ne 1/\sigma _{maj}$,
which will be important below.
The CIP conductivities are then
\begin{eqnarray}
\sigma ^P_{CIP,maj} &=& \sigma _{maj} \label{eq:sigmaPCIPmaj}\\
\sigma ^P_{CIP,min} &=& \sigma _{min} \label{eq:sigmaPCIPmin}\\
\sigma ^{AP}_{CIP,maj} = \sigma ^{AP}_{CIP,min} &=& \frac {\sigma
_{min}+\sigma _{maj}}2 = \frac{\sigma_{CIP}}2 .
                         \label{eq:sigmaAPCIP}
\end{eqnarray}

Within the two channel conduction model, for a given geometry and
magnetic orientation the conductivity of the sample is the sum of
the conductivities of the two spin channels. According to Eqs.\
(\ref{eq:sigmaPCIPmaj}) - (\ref{eq:sigmaAPCIP}), within this model
the parallel and antiparallel configurations in the CIP geometry
have the same sample conductivity, $\sigma_{CIP}$. Due to long
length scale interface disorder each of the conductivities in
Eqs.\ (\ref{eq:sigmaPCPPmaj}) - (\ref{eq:sigmaAPCIP}) changes
according to Eqs.\ (\ref{eq:alphadef}), (\ref{eq:alphaCIP}), and
(\ref{eq:alphaCPP}). The net conductivity and resistivity of the
sample therefore also change. As in the previous section we denote
the change in the resistivity by $\delta \rho = -\delta \sigma
/\sigma ^2$. Here, the resistivities are different for the two
geometries as well as the two magnetic orientations, so there are
a total of four $\delta \rho$'s, which are given below.
\begin{eqnarray}
\delta \rho ^P_{CPP} &=& 16\left( \frac A{\xi} \right)^2
    \bigg\{
    \frac {\rho _{min}^2( \sigma_{maj}^{-1} - \rho _{maj})}
         {(\rho _{min}+\rho_{maj})^2}
          \label{eq:deltarhoPCPP} \\
    &+&
    \frac {\rho _{maj}^2( \sigma_{min}^{-1} - \rho _{min})}
         {(\rho _{min}+\rho_{maj})^2} \bigg\}
          \nonumber \\
\delta \rho ^{AP}_{CPP} &=& 16\left( \frac A{\xi} \right)^2
             \bigg\{\frac 1{\sigma _{CIP}}
            -  \rho _{CPP}^{AP} \bigg\}
             \label{eq:deltarhoAPCPP} \\
\delta \rho ^P_{CIP} &=& 16\left( \frac A{\xi} \right)^2
            \frac 1{(\sigma _{CIP} )^2}
             \bigg\{ \sigma _{CIP}
               - \frac 1{\rho _{CPP}^P} \bigg\}
         \label{eq:deltarhoPCIP} \\
\delta \rho ^{AP}_{CIP} &=& 16\left( \frac A{\xi} \right)^2
            \frac 1{(\sigma _{CIP})^2}
             \bigg\{ \sigma _{CIP}
               - \frac 1{\rho _{CPP}^{AP}} \bigg\}
             \label{eq:deltarhoAPCIP}
\end{eqnarray}

Using the definitions of $\rho _{maj}$, $\rho _{min}$, $\sigma
_{maj}$, and $\sigma _{min}$, it follows that in the CPP geometry
the change in resistivity is negative, $\delta \rho _{maj/min}
<0$, while in the CIP geometry the change in the resistivity is
positive, $\delta \rho _{maj/min} > 0$.  This result is expected
from Figs.\ \ref{fig:CPPcurrent} and \ref{fig:CIPcurrent2}. In the
CPP geometry the waviness of the interfaces allows the current to
travel a greater distance through the less resistive layers,
thereby reducing the resistance. In the CIP geometry the waviness
of the interfaces disrupts the current flow through the low
resistivity layers, increasing the resistance.

The GMR is determined by the ratio of the parallel and
antiparallel resistivities.  As seen above, the parallel and
antiparallel resistivities either both increase or decrease
depending on the geometry.  Thus, one must compute their ratio
explicitly to determine whether the GMR increases or decreases
with long length scale interface disorder. Expanding Eq.\
(\ref{eq:GMR}) for the GMR to linear order in $\delta\rho_{P}$ and
$\delta\rho_{AP}$ and substituting the changes in the
resistivities of Eqs.\ (\ref{eq:deltarhoPCPP}) -
(\ref{eq:deltarhoAPCIP}), the change in the GMR for the CPP and
CIP geometries is
\begin{eqnarray}
\delta {GMR}_{CPP}
             &=& 16\left( \frac A{\xi} \right)^2
                 \bigg\{
                 \left(
                 \frac 1{\sigma _{CIP}\rho _{maj}} -
                 \frac {\rho _{CPP}^{AP}}{\sigma _{maj}\rho _{maj}^2}
                 \right)
                 \nonumber \\
             &+&
                 \left(
                 \frac 1{\sigma _{CIP}\rho _{min}} -
                 \frac {\rho _{CPP}^{AP}}{\sigma _{min}\rho _{min}^2}
                 \right)
                 \bigg\}
                 \label{eq:deltaGMRCPP} \\
\delta {GMR}_{CIP} &=& 16\left( \frac A{\xi} \right)^2
                 \frac 1{\sigma _{CIP}}
                 \left\{
                    \frac 1{\rho _{CPP}^P} -
                    \frac 1{\rho _{CPP}^{AP}}
                 \right\}.
                 \label{eq:deltaGMRCIP}
\end{eqnarray}
Using Eqs.\ (\ref{eq:deltaGMRCPP}) and (\ref{eq:deltaGMRCIP}), it
can be shown that in the short mean free path limit the GMR
decreases in the CPP geometry ($\delta GMR_{CPP} < 0$) and
increases in the CIP geometry ($\delta GMR_{CIP}> 0$).  In the CIP
case, this result is consistent with our numerical computation of
the GMR, which always increased with long length scale interface
roughness.  In the CPP case, however, we found numerically that
the GMR could either increase or decrease with interface roughness
depending on the sample geometry and the conductivities 
of the paramagnetic and ferromagnetic layers.  In both geometries, Eqs.\ 
(\ref{eq:deltaGMRCPP}) and (\ref{eq:deltaGMRCIP}) generally yield much
larger changes in the GMR than we predict numerically.  This
discrepancy is due to
the fact that the agreement between the analytic expressions for $\alpha$ 
and the numerical values varies depending on the layer conductivities.  
If the analytic and numeric values for $\alpha$ agree very well within
a particular spin channel, then the analytic expression for the change
in the conductivity for that channel will agree very well with the
change computed numerically.  When one considers a different spin
channel with different layer conductivities, the analytic $\alpha$
will generally be a worse approximation of the numeric value because the layer
conductivities have changed.  The change in
the conductivity for this channel computed analytically will 
likewise be a worse approximation of the numeric value.  This will
result in deviations between the antiparallel and parallel 
conductivities computed analytically and numerically.  
Since the GMR involves a
ratio of these quantities, the error will be more significant in the
GMR, 
resulting in large deviations between the analytic and numeric 
changes in the GMR.

\subsection{Estimates}

Both the long and short mean free path effects described in this
paper must be present to some extent in any sample.  The real
question is how large these effects are and whether they can
account for what is seen experimentally.  In this section we
estimate the size of the two effects in Fe(3nm)/Cr(1.2nm) multilayers 
using experimentally
determined parameters.  Although there are no adjustable
parameters in these calculations, they are still merely estimates
because the actual experiments are in neither the long nor the
short mean free path limits, but somewhere in between. This can
easily be seen from Cyrille \emph{et al.}'s data \cite{Schuller1}. In the long
mean free path limit the CPP and CIP resistances are the same,
whereas experimentally the ratio of the CPP to the CIP resistances
is roughly 1.5.  In the short mean free path limit there is no CIP
GMR for flat interfaces, while there is roughly a 10\% CIP GMR
observed experimentally. In addition to taking the long and short
mean free path limits, the calculations in this paper are
performed in two dimensions instead of three dimensions, and we
use one of the simpler Boltzmann equations.

To estimate the size of the long mean free path effect, we use
Eqs.\ (\ref{eq:Pexpansion}) and (\ref{eq:APexpansion}).  The
resistivities, $\rho _{min}$ and $\rho _{maj}$, may be determined
experimentally from Cyrille \emph{et al.}'s measurements of $\rho ^P$ and $\rho
^{AP}$ using Eqs.\ (\ref{eq:rhoP}) and (\ref{eq:rhoAP}). Although
the data vary with the number of bilayers, the results depend only
weakly on which data points one uses to compute $\rho_{maj}$ and
$\rho_{min}$, with the values in the range $\rho _{min}=45.7 \pm
0.9 \mu\Omega cm$ and $\rho _{maj}=130 \pm 8 \mu\Omega cm$. These
are very close to the values obtained from Zambano's data
\cite{Bass} of $\rho _{min}=45\mu\Omega cm$ and $\rho
_{maj}=143\mu\Omega cm$, which is not surprising because the
overall resistances in the two sets of experiments are close
together, even though the trends are not the same.

To get the changes in the majority and minority surface
resistances, $\delta \rho^*_{maj}$ and $\delta \rho^*_{min}$, one
needs to know the values of the surface resistances that the
changes will be computed from. These parameters have not been
determined in the experiments by Cyrille \emph{et al}.; however,
they have been determined in a series of experiments by Zambano
\emph{et al}.\ in which measurements were taken with different
layer thicknesses and number of layers \cite{Bass}. In their
notation $\rho _{maj}^*$ and $\rho _{min}^*$ are equal to
$AR^{\uparrow}_{Fe/Cr}$ and $AR^{\downarrow}_{Fe/Cr}$,
respectively, and they find these values to be $\rho _{maj}^* =
2.7 f\Omega m^2$ and $\rho _{min}^* = 0.5 f\Omega m^2$. We will
use these numbers for our estimate because the overall resistances
in the two sets of experiments are close to one another.

Finally, we need to determine the waviness or roughness of the
interfaces in the experiments.  This was quantified by Cyrille
\emph{et al}.\ using two techniques -- low angle x-ray diffraction
and energy filtered imaging using cross-sectioned samples in a
transmission electron microscope. \cite{Schuller1} Both techniques
show that the root mean square deviation of the layer height
increases within the multilayer and has the form $\sigma = \sigma
_a M^\alpha$, where $\sigma _a = 0.37$nm is the root mean square
deviation of the first bilayer, $M$ is the bilayer index, and the
exponent $\alpha$ equals 0.4. For the samples deposited at a
constant pressure of 5mTorr, the average distance between surface
bumps, or the period $\xi$ in our model, was determined to be
10nm, independent of the bilayer index. \cite{Schuller2} For a
multilayer with $N$ bilayers, the average value of $(A/\xi)^2$
used in Eq.\ (\ref{eq:longmfp3}) to compute the changes in the
interface resistances is thus
\begin{equation}
\langle ( A/{\xi}) ^2 \rangle =
\frac 1N \sum _{M=1}^N (\sqrt{2}\sigma _a/\xi )^2
M^{2\alpha},
\label{eq:avgroughness}
\end{equation}
where we have used the fact that the root mean square fluctuation
for a sine wave is $A/\sqrt {2}$.

\begin{figure}
  {\resizebox{7.5cm}{!}{\includegraphics{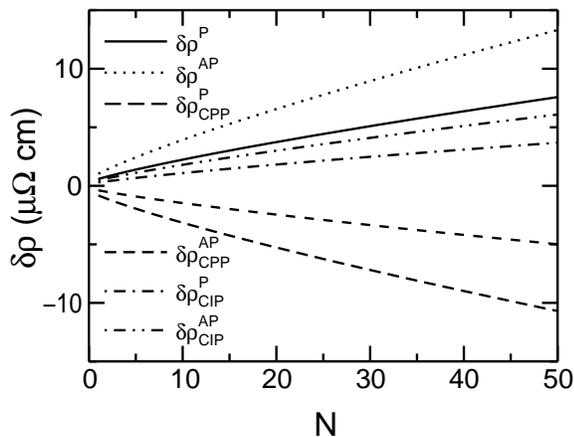}}}
  \caption{\label{fig:estimate} Estimates in the long and 
  short mean free path limits
  for the changes in the
parallel and antiparallel resistivities due to long length scale
interface roughness. The expressions for the changes in the
resistivities are evaluated using experimentally determined
parameters, but the data shown are only estimates because the
experiments are in between the long and short mean free path
limits. The magnitudes of the resistivity changes are consistent
with those measured by Cyrille \emph{et al}.; however, for
intermediate mean free paths there can be cancellations of $\delta
\rho ^{P/AP}$ and $\delta \rho ^{P/AP}_{CPP}$, which may explain
the results of Zambano \emph{et al}. For the CIP geometry there is
no such cancellation and the estimates are comparable to the
changes seen by Cyrille \emph{et al}. }
\end{figure}

The result of using Eqs.\ (\ref{eq:Pexpansion}),
(\ref{eq:APexpansion}), (\ref{eq:longmfp3}), and
(\ref{eq:avgroughness}) to estimate the size of the long mean free
path effect is shown in Fig.\ \ref{fig:estimate}. As expected both
the parallel resistivity (solid line) and the antiparallel
resistivity (dotted line) increase with roughness. The size of the
increase in the antiparallel resistivity measured in the
experiments by Cyrille \emph{et al}.\ is comparable to this
estimate for the antiparallel resistivity.  However, our estimate
of the size of the increase in the parallel resistivity is much
larger than the negligible or possibly negative change in the
parallel resistivity seen in the same experiments.

In the short mean free path limit, there are four resistivities
for the two geometries (CIP and CPP) and magnetic configurations
(P and AP). The changes in the resistivities for these four cases
are given in Eqs.\ (\ref{eq:deltarhoPCPP}) -
(\ref{eq:deltarhoAPCIP}). In evaluating these equations, we use
the CPP resistivities, $\rho ^P_{CPP}$ and $\rho ^{AP}_{CPP}$,
determined by the same $\rho _{maj}$ and $\rho _{min}$ used in the
long mean free path limit using Eqs.\ (\ref{eq:rhoP}) and
(\ref{eq:rhoAP}). The CIP conductivity , $\sigma _{CIP}$, is taken
to be in between the observed parallel and antiparallel
conductivities: $(\sigma _{CIP})^{-1} = 25 \mu\Omega cm$. Finally,
the conductivities $\sigma _{maj}$ and $\sigma _{min}$ are not
readily determined from experiment.  To estimate them, we impose
the condition $\sigma _{maj}/\sigma _{min} = \rho _{min}/\rho
_{maj}$ and use the value for $\sigma _{CIP} = \sigma _{min} +
\sigma _{maj}$ from above.
The resulting changes in the resistivities are shown in Fig.\
\ref{fig:estimate}. As expected, here the changes in the
resistivities for the CPP configuration are negative, while the
changes in the resistivities for the CIP configuration are
positive.  Although it was mentioned previously that the 
analytic expressions for
the changes in the GMR given in Eqs.\ (\ref{eq:deltaGMRCPP}) and 
(\ref{eq:deltaGMRCIP}) predict a much larger effect in the short 
mean free path limit than we predicted numerically, the
changes in the resistivities presented here are more reliable.  Since
the GMR involves a ratio of the antiparallel and parallel
resistivities, the error in the analytic GMR will be more significant
than the error in the analytic antiparallel and parallel resistivities.

As Fig.\ \ref{fig:estimate} illustrates, the changes in the CPP
resistivities are of roughly the same magnitude, but opposite
sign, in the long and short mean free path estimates. Thus, in a
sample with intermediate mean free paths, there may be substantial
cancellation of these two effects.  For the CIP configuration,
however, both the long and short mean free path effects tend to
increase the resistivity by a similar magnitude. An increase in
the CIP resistivity of comparable size to these estimates is
indeed seen in the experiments by Cyrille \emph{et al}.

These estimates show that the effects described in this paper are
of the correct size to describe the observations of Cyrille
\emph{et al}.; however, detailed quantitative comparison is not
possible because the mean free paths in the experiments are in
between the long and short mean free path limits.  It is entirely
possible that for intermediate mean free paths the two effects
cancel, leading to the negligible change in the resistivity seen
in the experiments by Zambano \emph{et al}.  If the long mean free
path effect does dominate, then a result similar to that of
Cyrille \emph{et al}.\ would be observed.  In the CIP
configuration there are no complications due to the effects
canceling. The estimated change in the resistivity in both limits
is positive and comparable to the increase observed by Cyrille
\emph{et al}.

\section{Conclusions}

In this paper we have examined the effect of interface disorder
which is long on the atomic scale. These kinds of fluctuations are
ubiquitous in metallic multilayers. A semiclassical Boltzmann
equation was solved in both the limits where the electronic mean
free paths were short and long compared to the layer thicknesses.

In the short mean free path case the current flow is nonuniform,
and long length scale interface disorder increases the effective
conductivity in the CPP geometry and decreases the effective
conductivity in the CIP geometry.  In the CPP case, the effect is
due to interface disorder providing a less-resistive, non-linear
path of current flow.  In the CIP case, the effect results from
interface roughness disrupting the flow of current through low
resistance layers.  In the long mean free path case the current
flow is uniform in the CPP and CIP geometries. The resistance
increases with long length scale interface roughness in both
geometries because of the additional scattering created by longer
interfaces for the disordered layers than the flat layers.

The experiments discussed in this paper are in neither the short
nor the long mean free path limits, but somewhere in between.
Nonetheless, in estimating the size of the long and short mean
free path effects we find that in the CPP geometry the increase in
the antiparallel resistivity observed by Cyrille \emph{et al}.\ is
comparable to the increase we predict in the long mean free path
limit. The increase estimated in the CPP geometry for the parallel
resistivity, however, is larger than the negligible increase
measured by Cyrille \emph{et al}.  We find that the long and short
mean free path effects tend to cancel for intermediate mean free
paths in the CPP geometry, which could explain the observations of
Zambano \emph{et al}.  For the CIP geometry, in both the long and
short mean free path limits we estimate an increase in the
resistivity that is comparable in magnitude to the increase seen
by Cyrille \emph{et al}. Therefore, the effects described here may
be the source of the experimentally observed increase in the
resistivity with long length scale disorder, although clearly more
theoretical investigation is needed to understand the crossover
between the long and short mean free path limits.

\begin{acknowledgments}
The authors would like to thank Tat-Sang Choy, Jack Bass, and Ivan
Schuller for helpful discussions. This research was supported by
DOD/AFOSR Grant No. F49620-96-1-0026, the Center for Condensed
Matter Sciences, the University Scholars Program at the University
of Florida, and the National Science Foundation through the U.F.
Physics REU Program.
\end{acknowledgments}

\end{document}